\documentclass[useAMS,usenatbib]{mn2e}

\usepackage{psfig, epsf, epsfig}

\title[Globular cluster formation]{Globular cluster formation
with multiple stellar populations: A single-binary 
composite scenario}
\author[K. Bekki]
{Kenji Bekki${}^1$\thanks{E-mail:
kenji.bekki@uwa.edu.au} \\
${}^1$ICRAR M468
The University of Western Australia
35 Stirling Hwy, Crawley
Western Australia 6009, Australia}

\begin{document}

\date{Accepted, Received 2005 February 20; in original form }

\pagerange{\pageref{firstpage}--\pageref{lastpage}} \pubyear{2005}

\maketitle

\label{firstpage}

\begin{abstract}

We discuss a GC formation scenario in which
the first generation (1G) of single 
asymptotic giant branch (AGB) stars
and intermediate-mass close binaries (IMCBs)
eject gas, from which the second generation (2G) of  stars can be formed.
The two key parameters 
in the scenario are  the fractions of binary stars ($f_{\rm b}$)
and the slopes
($\alpha$) of the stellar initial mass functions (IMFs) for 1G stars.
Principle results derived by analytic and
one-zone models of GC formation are as follows.
The mass fraction of 2G stars ($f_{\rm 2g}$)  can be higher than
$\approx  0.4$ for 
$\alpha < 1.8$ and  is not so dependent on $f_{\rm b}$.
The ratio of the initial mass of a  GC to the present-day mass ($M_{\rm gc}$)
ranges from 2 to 7 depending on $\alpha$
for $0.5 \le f_{\rm b} \le 0.9$.
The differences in [Na/Fe] between 1G and 2G stars can
be as large as 0.7 for a wide range of model parameters.
The Li abundances of 2G stars can be as high as those of 1G
even if the pristine gas from IMCBs is assumed to be Li-free.
Formation histories  of 2G stars show at least two peaks owing to 
two peaks in the total ejection rate of gas from IMCB populations.
The observed correlation between $f_{\rm 2g}$ and $M_{\rm gc}$
can be due to $\alpha$ depending on $M_{\rm gc}$.
The hypothetical long duration of
2G formation  ($\approx 10^8$ yr) is possible,
because 
massive star formation 
can be suppressed through frequent
dynamical interaction between 1G stars and gas clouds.

\end{abstract}


\section{Introduction}

Most Galactic globular clusters (GCs) are observed to have chemical abundance spreads
among their individual stars (see recent reviews by Gratton et al. 2019, G19; 
Milone and Marino  2022; MM22). This observed presence of  multiple populations (MPs) 
in GCs is a fundamental GC characteristic that any theory of GC formation needs to
explain.  The MP phenomena are quite diverse, including
large helium abundance spread in $\omega$ Cen (e.g., Piotto et al. 2005),
anti-correlations 
between the chemical abundances of light elements (e.g., Carretta et al. 2009),
Type I and II dichotomy (e.g., Marino et al. 2017),
C+N+O abundance spreads (e.g., Yong et al. 2012) 
abundance spread in $s$-process (Marino et al. 2015) 
and $r$-process elements
( e.g., Roederer \& Snedin 2011 for M15),
and large age and [Fe/H] differences in Terzan 5 (e.g., Ferraro et al. 2009).
These observed properties of GCs can provide useful constraints on the theory of
GC formation, however,
previous theoretical models of GC formation with MPs appear to have potentially
serious problems in explaining all of these
in a self-consistent manner (e.g., Renzini et al. 2015;  Bastian \& Lardo 2018, BL18).

Most previous theoretical models of GC formation with MPs assumed
that a GC  consists of two major populations, i.e.,
the first generation (``1G'') of stars formed within natal
gas clouds, and 
the second  generation  (``2G'')  of stars that
are formed from 
gaseous ejecta from 1G stars mixed with (or ``diluted'' by)
``pristine'' gas that have the same chemical abundances as
those of natal clouds (i.e., 1G stars).
The origin of pristine gas and the dilution processes
are the  two key problems in previous theoretical models of GC formation,
and different models adopt different assumptions to solve the two
(see BL18 for critical reviews
of these models).
One of the most extensively investigated models is 
the so-called  ``AGB scenario'' in which 2G stars form  from AGB ejecta
diluted by interstellar medium (ISM) of GC-hosting gas-rich
galaxies (e.g.,  Fenner et al., 2004;
Bekki et al. 2007;  D'Erocle et al. 2008, D08).
Although Ventura et al. (2001) first proposed that low-mass
stars can accrete AGB ejecta to show lower [O/Fe] and higher 
[Na/Fe] and [Al/Fe], their scenario did not assume dilution by ISM.

\begin{table}
\centering
\begin{minipage}{90mm}
\caption{Physical meanings of acronyms and
model parameters for globular clusters (GCs).}
\begin{tabular}{ll}
Acronym & Physical meaning  \\
IMCB & Intermediate-mass close binary  \\
RLOF & Roche lobe overflow in binary stars  \\
MP & Multiple stellar population  \\
IMF & Initial mass function of stars  \\
1G & First generation of stars  \\
2G & Second generation of stars  \\
nG & n-th  generation of stars (n=1,2,3,...)  \\
$m$ & Mass of an individual star (${\rm M}_{\odot}$) \\
$m_{\rm w}$ &  Mass of AGB wind (${\rm M}_{\odot}$) \\
$q$ & Mass ratio of binary stars \\
$\alpha$ & IMF slope for 1G stars \\
$m_{\rm l}$ & Lower mass cut-off for the IMFs of all stars \\
$m_{\rm u, 1g}$ & Upper mass cut-off for the IMF of 1G stars\\
$m_{\rm u, 2g}$ & Upper mass cut-off for the IMF of 2G stars\\
$M_{\rm i}$ &  Initial total  mass of a GC with 
$m_{\rm l}\le m \le m_{\rm u,1g}$\\
$M_{\rm gc}$ & Total mass of a GC with 
$m_{\rm l}\le m \le 0.8 {\rm M}_{\odot}$\\
$M_{\rm gc, g}$ & Total gas mass of a GC  \\
$M_{\rm 1g}$ & Total  mass of 1G stars with 
$m_{\rm l}\le m \le 0.8 {\rm M}_{\odot}$\\
$M_{\rm 2g}$ & Total  mass of 2G stars with 
$m_{\rm l}\le m \le 0.8 {\rm M}_{\odot}$\\
$f_{\rm 2g}$ & Mass fraction of 2G stars ($=M_{\rm 2g}/M_{\rm gc}$) \\
$f_{\rm b}$ & Mass fraction of binary stars in a GC \\
$f_{\rm imcb}$ & IMCB fraction among intermediate-mass binaries  \\
$F_{\rm mb}$ & Mass budget factor ($=M_{\rm i}/M_{\rm gc}$) \\
$M_{\rm g,s}$ & Total mass of ejecta from single AGB stars \\
$M_{\rm g,b}$ & Total mass of ejecta from IMCBs \\
$M_{\rm g}$ & Total gas mass from single AGB and IMCB stars \\
{\it \. M} & Gas accretion rate in a GC \\
$f_{\rm ej}$ & Mass fraction of ejecta from IMCBs   \\
$f_{\rm pr}$ & Mass fraction of pristine gas in the ejecta of IMCBs \\
$F_{\rm dil}$ & Dilution factor
($=f_{\rm pr} M_{\rm g,b}/M_{\rm g}$) \\
$f_{\rm g, b}$ & Fraction of gas from IMCBs among all gas  \\
$\Delta{\rm [Na/Fe]}$  & Difference in [Na/Fe] between 1G and 2G stars \\
${\rm [Na/Fe]}_{\rm min}$  & Minimum [Na/Fe] in 1G stars \\
${\rm [Na/Fe]}_{\rm max}$  & Maximum [Na/Fe] in 2G stars \\
A(Li)  & Li abundance (in 2G stars) \\
A$_{\rm m}$(Li)  & Mass-weighted  Li abundance of AGB ejecta \\
$Y_{\rm Li}(m)$  & Li fraction from  AGB stars depending on $m$ \\
$Y_{\rm Li, m}$  & Mean Li fraction among all AGB ejecta  \\
$\beta$  & Power-law slope in the adopted SF law \\
$M_{\rm g,th}$  & Threshold gas mass for star formation \\
$\epsilon_{\rm sf}$  &  Star formation efficiency of 2G stars  \\
\end{tabular}
\end{minipage}
\end{table}

One of potential problems in the standard AGB scenario is the origin of
pristine gas (e.g., D'Ercole et al. 2011). Interstellar medium (ISM)
of galaxies hosting GCs is often assumed to be pristine gas that 
can be  mixed 
with AGB ejecta and subsequently converted into 2G stars
(e.g., Bekki et al. 2007; D'Ercole et al. 2010;
Calura et al. 2019;
McKenzie \& Bekki 2021, MB21; Yaghoobi et al. 2022).
However, this assumption  of dilution of AGB ejecta
 by ISM has the following potential
problems. 
First, ISM needs to be accreted onto GCs to mix well with
AGB ejecta (and to form high-density regions for star formation)
when gas from
massive AGB stars  
is accreting onto GCs.
It is not so obvious that this requirement
can be met in all  GCs interacting with ISM in different
GC host galaxies.
As shown in MB21,  the processes of ISM accretion onto GCs
depends on a number of parameters such as the sizes
of gaseous ``holes'' in ISM surrounding GCs 
generated by multiple core-collapsed supernovae (CCSNe) of 1G populations 
and the mass density of ISM.
For example, if GCs are embedded in giant gaseous holes,
then ISM accretion onto GCs can occur later than the commencement
of AGB phases of intermediate-mass stars:
no dilution is possible and new stars can be formed
from pure AGB ejecta.
This problem is refereed to as the ``timing'' problem in the present
study.

Second a right amount of ISM needs to be accreted onto GCs to explain
the observed mass fraction of 2G stars ($f_{\rm 2g}$)  and the abundance
differences in light elements  between 1G and 2G (``amount''
problem).
The observed mass fraction of 2G stars for GCs with $\log M_{\rm gc}=5.2$
is 0.4 (G19), which requires a certain range of ISM mass. Too much accretion
of ISM onto GCs can end up with large $f_{\rm 2g}$ that is not observed
and too small differences in light elements between 1G and 2G  due to dramatic dilution
of AGB ejecta by pristine ISM. Accordingly,  fine-tuning is required for the right
amount of accreted ISM onto GCs, though not only the physical properties of GCs
but also those of ISM (i.e., relative velocities between ISM and GCs)
can determine the total amount of ISM accreted onto GCs.

Third, the metallicity of ISM needs to be very similar to those of 1G
stars in GCs (``metallicity'' problem).
Recent observational studies have revealed that
Type I GCs show similar  [Fe/H] between 1G and 2G, though
each of the two populations exhibits [Fe/H] variations
(e.g., G19; Legnardi et al. 2022).
ISM can originate from different regions of GC host galaxies
(e.g., MB21), and CCSNe of 1G stars can pollute the ISM of the host dwarfs
to significantly increase the metallicities of ISM. 
Therefore, it is possible that [Fe/H] of ISM can be significantly different 
from 1G stars.
Fourth, accretion of ISM needs to stop before  stars with their original
masses lower than $3M_{\odot}$ enter into their AGB phases (``truncation'' problem).
This is because such AGB ejecta can have enhanced C+N+O
abundances (e.g., Fenner et al. 2004) and be mixed with ISM, which can 
lead to 2G formation
with their C+N+O abundances significantly higher than those of 1G stars.
These four (potential) problems related to dilution of AGB ejecta need to be
solved in GC formation  models based on single AGB stars.
It should be noted here that Renzini et al. (2022) also pointed out possible
problems of the standard AGB scenario.

Vanbeveren et al. (2012, V12) proposed an alternative model in which 
intermediate-mass close binaries (IMCBs) can eject  fresh pristine gas
that can be mixed with stellar winds of AGB stars and finally converted
into 2G stars.
In this model,  a significant amount of gas from 
IMCBs can be ejected from IMCBs during the Roche lobe overflow (RLOF) 
and the common envelope phases,
and some of the ejected gas can have chemical abundances that
are almost identical to those of 1G stars (i.e., pristine gas).
Accordingly, they suggested that if these pristine gas is mixed well
with stellar winds from single AGB stars, then
2G stars with the observed chemical abundance patters can be formed.
Although the idea of gas from binary stars in V12 is essentially
the same as the underling assumption
in ``massive interacting binary'' scenario by de Mink et al. (2009),
V12's scenario 
has not been discussed extensively so far: this scenario is referred to as
the ``SBC'' (single-binary composite) scenario just for convenience

Recent observational studies have shown that the {\it initial} fractions
of binary pairs of intermediate-mass stars is almost 100\%  in stellar associations
(e.g., Kouwenhoven et al. 2007, K07), though the binary fractions of the present-day GCs
are not so high (e.g., Sollima et al.  2007).
Also, recent numerical simulations of binary evolution 
in GCs have shown that the binary fractions can decrease by only 10\%
within one  dynamical relaxation time at half-mass radii corresponding to 
roughly $0.1$ Gyr (e.g., Fig. 1 in Hong et al. 2015).
Accordingly, GCs should have rather high fractions of binary stars
$\approx 0.1$ Gyr after their formation, when a large amount
of AGB ejecta of 1G stars are accreting onto their  central regions.
Thus the SBC scenario needs to be investigated  by
theoretical models of GC formation not only because it can possibly
solve the problems related to dilution of AGB ejecta mentioned above,
but also because binary stars dominate $\approx 0.1$ Gyr old GCs
so that they can influences the formation of 2G stars.

The purpose of this paper is thus  to discuss the basic characteristics of
the SBC scenario using simple analytic calculations and one-zone models
for the physical properties of GCs with MPs. 
In discussing the MP problems of GCs,
we adopt a bold assumption that GCs do not accrete gas from ISM of their
host galaxies in order to more clearly understand the basic characteristics
of the SBC scenario: we here admit that such an assumption could be rather
idealized, given that recent hydrodynamical simulations demonstrated
that ISM accretion onto GCs is possible (e.g., McKenzie \& Bekki 2018;  MB21).
We focus exclusively on the global properties of GCs predicted from
the SBC scenario,  such as
 mass fractions of 2G stars, [Na/Fe] differences
between 1G and 2G stars,  star formation histories in GCs.
Accordingly we will discuss
their  3D structures and kinematics, detailed chemical abundance patterns,
and mass-dependences of these 
using  hydrodynamical
simulations of GC formation in our forthcoming papers.

The plan of the paper is as follows.
We outline the SBC scenario and possible advantages and disadvantages
of it in explaining the observed properties of GCs
in \S 2.
We describe  simple analytic models for the scenario
and present the key results 
in \S 3.
Based on one-zone models,
we investigate  the possible star formation histories of 2G populations in GCs
in \S 4.
We discuss a number of key problems related to the formation of MP in GCs 
based on the present new results.
in \S 5.
We summarize the key characteristics of the scenario 
in \S 6.

In this paper, we do not discuss feedback effects of various evolved stellar
populations and pulsar winds, which have been investigated by several authors
(e.g., Naiman et al. 2020). These feedback effects could remove some of the intra-cluster
gas ejected from massive AGB stars and binary stars to end up with much less
efficient star formation of 2G stars. 
We do not discuss the possible anti-correlations between light elements 
(e.g., Na-O anti-correlation)
in the present model either, simply because the present study does not include
chemical evolution at all. 
Galaxy-scale physics related to GC formation with
multiple stellar populations such as the formation
of giant molecular clouds hosting young clusters and cloud-cloud
collisions in galaxies are also totally ignored in the present study,
though they are included in our previous simulations on cluster formation
(e.g., Bekki et al. 2004; Williams et al. 2021).
We will discuss these key issues in our future works using more sophisticated
hydrodynamical simulations of GC formation.

\section{The SBC scenario}

\subsection{Outline}

In this SBC scenario,  massive compact stellar systems consisting
of 1G stars are first formed within their host gas-rich dwarf
disk galaxies with high mass densities
owing to dynamical instabilities of the disks (B19a).
The GCs initially have rather high binary fractions
(almost 100\%) 
in intermediate-mass stars, as observed
in local star-forming regions  (e.g., K07),
and thus the binary populations
can start to eject gas through RLOF about $\sim 30$Myr after the initial
burst of 1G star formation. These ``pristine'' gas is accumulated
into the central regions of the GCs so that they can form high-gas
density regions within the deep gravitational potentials.
Then stellar winds from single AGB stars with 
$m \approx 8 {\rm M}_{\odot}$ start to be accumulated into the 
central regions to  mix with the pristine gas from IMCBs.
Finally new stars with peculiar abundance patterns
(e.g.,  Na-rich, O-poor) are formed from the mixed
gas very efficiency to develop the 2G population.

In the standard AGB scenario based on single AGB stars only,
the anti-correlation between CNO abundances can be explained
in the context of dilution of AGB ejecta with pristine gas (e.g., D'Ercole et al. 2010).
However, the origin of pristine gas between the AGB and SBC scenarios
is quite different. The amount of pristine gas and the epoch
when the gas can be accreted onto the central regions of 
pre-existing 1G populations are determined primarily by
the evolution of IMCBs within GCs in the SBC scenario.
This idea of pristine gas originating from
1G stars is essentially the same as the proposal by
Gratton \& Carretta (2010) in which
less evolved 1G stars can eject a significant amount of pristine gas
(though the amount is only at most 1\% of 1G stars).
Given that the amount of ejecta from single AGB stars can be 
determined by the properties of 1G stars in the SBC scenario,
the properties of 2G stars can be also determined largely by
GC properties themselves: it should be noted here that
the accretion rate  of pristine ISM onto GCs
 in the AGB scenario depends both on GC masses and  on
the properties of GC host galaxies.

Secondary star formation within GCs can continue until 
fresh supply of gas due to 
gas accretion from single AGB stars and IMCBs onto their 
deep potential wells is severely reduced. 
During this 2G formation,  the maximum masses of new stars
cannot be larger than $8 {\rm M}_{\odot}$ so that
feedback effects of CCSNe cannot influence the 2G formation
at least $\approx 300$ Myr: the possible physical reasons for this
are discussed later in this paper.
It should be noted here that  no observations have so far found
direct evidence of secondary star formation in young massive clusters
within nearby galaxies (e.g., BL18). However, it is possible that
proto-GCs could be quite different from these younger ones in their total
masses and environments, and such difference could be responsible for 
the origins of multiple stellar populations with different chemical
abundances.
Accordingly, the apparent lack of observational evidence for ongoing star formation
within young clusters with ages ranging from 30 to 300 Myr 
does not necessarily rule out the assumed secondary star formation
in young GCs.

In our previous numerical simulations of GC formation,
1G and 2G stars have been demonstrated to be formed
in gas-rich dwarf galaxies (e.g., B19a, MB21).
However, these simulations did not investigate at all
the time evolution of gas ejected from IMCBs within
GCs. Therefore, it is totally unknown (i) how much gas can be ejected
from single and binary intermediate-mass stars,
(ii) what the possible differences in chemical abundance patterns are between
1G and 2G stars,
and (iii) whether or not
the observed fundamental  properties of GCs with MPs cab be possibly explained.
It is thus the main purpose of this paper to provide basic predictions of the
scenario using rather idealized analytic and one-zone models of GC formation.

Gas ejected from AGB stars with different masses itself does not show 
the observed anti-correlation between light elements (e.g., [Na/Fe] vs [O/Fe]).
However, if Na-rich and O-poor gas
ejected from more massive AGB stars mix well with pristine Na-normal and O-normal
gas, then 2G stars formed from such mixed gas can show a Na-O anti-correlation.
We are currently investigating the anti-correlations between light
elements (CNO etc) for the simulated GCs and will discuss in what physical
conditions the observed Na-O anti-correlation can be reproduced well in the SBC
scenario (Bekki 2022).  The preliminary results suggest that
the timescale of 2G formation should be less than $10^8$ yr in order to
reproduce the Na-O anti-correlation: if gas from low-mass AGB stars is converted
into new stars, then the observed anti-correlation cannot be reproduced.

\begin{figure*}
\psfig{file=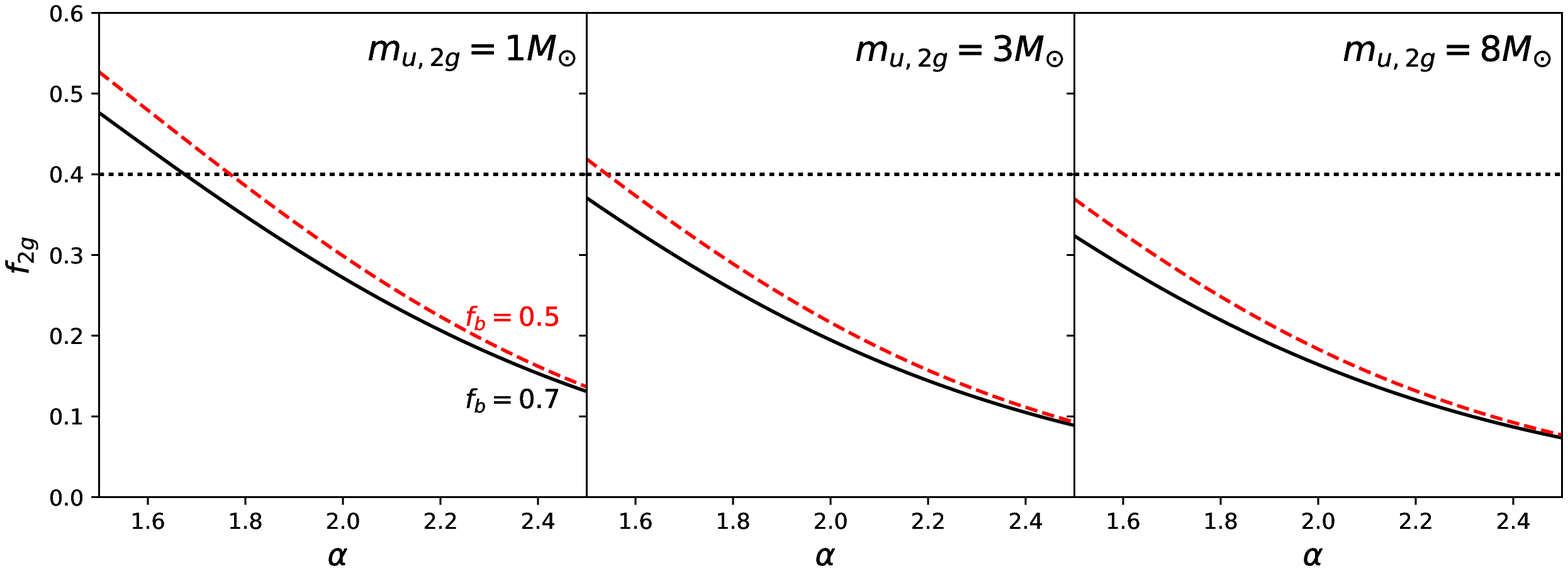,width=18.0cm}
\caption{
Dependence of $f_{\rm 2g}$ (2G mass fraction) on $\alpha$ (IMF slope)
for $m_{\rm u, 2g}=1 {\rm M}_{\odot}$ (left),
$3 {\rm M}_{\odot}$ (middle)
and $8 {\rm M}_{\odot}$ (right). Black solid and red dashed line
indicate the models with $f_{\rm b}=0.7$ and 0.5,
respectively, and the observed $f_{\rm 2g}=0.4$
is shown by a horizontal dashed line in each frame.
}
\label{Figure. 1}
\end{figure*}

\subsection{Several possible advantages of the scenario}

There are the  following advantages in this scenario in 
explaining the origin of GCs with multiple stellar populations.
First,
in this scenario, pristine gas from IMCBs 
can be ejected through RLOF
and rapidly accumulated within the deep potential
wells of proto GCs well before single massive
AGB stars with $m \approx 8 {\rm M}_{\odot}$
eject polluted gas. Therefore, ejecta from
such AGB stars can be mixed with the pristine gas
and then converted into new stars, if the physical conditions
for star formation can be met for the mixed gas.
Thus, there is no ``timing problem'' (in the required dilution processes)
in this scenario.

Second, 
the total amount of pristine and polluted gas can be determined
by the physical properties of GCs themselves, such as
number fraction of binary stars
($f_{\rm b}$),
mass fraction of pristine gas ejected from binary stars ($f_{\rm pr}$),
and initial total masses of  GCs ($M_{\rm i}$): there is a possibility
that the ``right amount'' of pristine gas can be supplied from IMCBs through
RLOF for a reasonable set of the above parameters.
Therefore, the physical properties
of GCs with multiple stellar populations,
such as fractions of 2G stars ($f_{\rm 2g}$)
and the mass budget factor ($F_{\rm mb}$, i.e., the ratio
of initial to present-day GC mass, defined later in detail), which
is the ratio of the initial total mass of a GC to the present-day
mass ($M_{\rm i}/M_{\rm gc}$),
can be determined by GC properties themselves.

Third, the required
cessation of star formation in 2G populations of GCs  can be naturally
explained in this scenario (no ``truncation'' problem): 
here secondary
star formation needs to be avoided because of the formation of 2G stars
with non-constant C+N+O from ejecta of low-mass AGB stars
 (e.g., Fenner et al. 2004).
Fresh supply of gas from IMCBs for secondary star formation can be 
severely suppressed within $\approx 300$ Myr  after GC formation,
because the ejection rate of pristine gas from IMCBs become
very low after $\approx 300$ Myr (see Fig. 2 in V12).
Such suppression of gas supply is highly likely
to end up with no/little further star formation owing to low gas
densities within GCs.  
Fourthly, the pristine gas from IMCBs have the same [Fe/H]
as the polluted gas from single AGB stars in a GC, which means that
[Fe/H] of 2G stars should be the same as their 1G counterparts
(no ``metallicity'' problem).

Fifth, 2G formation can occur only after all CCSNe are exploded.
In the massive binary scenario by de Mink et al. (2009),
2G stars need to be formed 
sometime between (i)  after  the ejection of gas from massive binary stars
and (ii) before massive CCSNe: only less than $\approx 3$ Myr is allowed for
the gas from the binary stars to be cooled down and subsequently converted into
new stars. This problem of very narrow time window for 2G formation
does not exist in the SBC scenario.
These advantages imply that the SBC scenario is very promising,
however, the scenario has several possible problems in reproducing
the observed properties of GCs, as we discuss later.
Also, we need to discuss the SBC scenario in the context
of well known problems
of GC formation, e.g., the observed $f_{\rm 2g}$ depending on
$M_{\rm gc}$, the so-called mass budget problem, 
difference in Li  abundances between 1G and 2G stars etc
(e.g., G19, MM22)
in a quantitative way.

\begin{figure}
\psfig{file=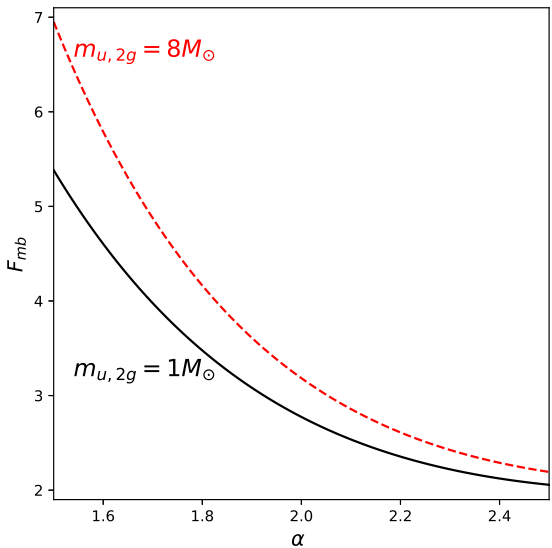,width=8.5cm}
\caption{
Dependence of $F_{\rm mb}$ (mass budget factor) on $\alpha$ 
in the models with
$m_{\rm u, 2g}=1 {\rm M}_{\odot}$ (black solid) and
$m_{\rm u, 2g}=8 {\rm M}_{\odot}$ (red dashed).
Since this dependence of $F_{\rm mb}$ on $\alpha$ is very similar 
between different $f_{\rm b}$,
only the models with $f_{\rm b}=0.7$ are shown.
}
\label{Figure. 2}
\end{figure}

\section{Basic characteristics of the scenario}
\subsection{Analytic models}

Using analytic models, we discuss how $f_{\rm 2G}$,
$F_{\rm mb}$, Li abundances of 2G stars (A(Li)),
and [Na/Fe] differences between 1G and 2G stars depend
on the model parameters of the SBC scenario.
Since it is assumed that a GC consists of 1G and 2G stars,
the present-day total mass of a GC ($M_{\rm gc}$) is  as follows:
\begin{equation}
M_{\rm gc} = M_{\rm 1g} + M_{\rm 2g}.
\end{equation}
The initial total mass of the GC ($M_{\rm i}$) consisting only
of 1G stars
 is significantly larger
than $M_{\rm gc}$ owing to the stellar mass loss through
supernovae, stellar winds, internal dynamical evolution,
and tidal stripping. 
The 1G stars can lose their stellar masses significantly through
stellar evolution only  (e.g., 
17\% from SNe for a canonical IMF), and they can lose much masses
due to internal dynamical evolution and tidal stripping if they have more diffuse
spatial distributions compared to 2G stars (e.g., Vesperini et al. 2010).
The mass budget factor
($F_{\rm mb}$)  is defined as follows:
\begin{equation}
F_{\rm mb}= \frac{ M_{\rm i} }{ M_{\rm  gc} } .
\end{equation}

The adopted power-law IMF in number for each GC is  defined 
as follows;
\begin{equation}
\psi (m) = C_{0}m^{-\alpha},
\end{equation}
where $m$ is the initial mass of
each individual star and the slope $\alpha =2.35$
corresponds to the Salpeter IMF.
The normalization factor $C_0$ is a function of 
the initial mass of a GC
($M_{\rm i}$),
$m_{\rm l}$ (lower mass cut-off), and $m_{\rm u}$ (upper mass cut-off):
\begin{equation}
C_{0}=\frac{M_{\rm i}
\times (2-\alpha)}{{m_{\rm u}}^{2-\alpha}-{m_{\rm l}}^{2-\alpha}}.
\end{equation}
Although we use $\alpha=2.3$ (Salpeter IMF)
for 2G stars,
we consider that $\alpha$ is a free parameter in the present study.
We also consider that $m_{\rm u}$ can be different between
the two populations: it is defined as $m_{\rm u, 1g}$
and $m_{\rm u, 2g}$ for 1G and 2G stars, respectively.
Both 1G and 2G stars
have $m_{\rm l}$  ($=0.25 {\rm m}_{\odot}$),
and $m_{\rm u, 1g}$ is fixed at $50 {\rm M}_{\odot}$.

The mass fraction of 2G stars is defined as follows:
\begin{equation}
f_{\rm 2g}=\frac{ M_{\rm 2g} }{ M_{\rm gc} },
\end{equation}
where $M_{\rm 2g}$ and $M_{\rm gc}$ are  estimated for low-mass stars
with $0.25 \le m/{\rm M}_{\odot} \le 0.8$.
Given that $\alpha$ and $m_{\rm l, 2g}$ are  fixed for 2G stars,
$f_{\rm 2g}$ is determined by  $\alpha$ and $m_{\rm u, 2g}$.
We also estimate the initial
total masses  of single and binary  stars ($M_{\rm i, s}$
and $M_{\rm i, b}$, respectively) and thereby
model the time evolution of the two populations separately.
We consider that the fraction of binary stars is the key parameter,
which is defined as follows:
\begin{equation}
f_{\rm b}=\frac{ M_{\rm i, b} }{ M_{\rm i} }.
\end{equation}

Using the observed properties of very young binary populations in Scorpius OB2 
associations, K07 revealed that the {\it current}
binary fraction
of A- and B-type stars is at least 70\% and also suggested that
the primordial fraction can be almost 100\%:
see Kroupa and Jerabkova (2018) for 
a recent review on the binary fractions depending
on stellar masses. If the distribution of orbital 
periods ($P$)
of binary stars is described as $f_{\rm p}(P) \propto P^{-1}$ with
the minimum and maximum $P$ being 0.5 days and 0.15 Myr (K07) respectively,
then the (original) fraction of IMCBs ($f_{\rm imcb}$) 
 among all binary populations with $P$ less than 4000
days is 0.71: the minimum
and maximum $P$  adopted in V12 are 1.0 and 3700 days, respectively.
If the maximum  $P$ of 10000 days is adopted for IMCBs that can lose
gas from the systems,
then $f_{\rm imcb}=0.8$.
Although $f_{\rm imcb}$  in GC formation could be
different from those of local OB associations
we consider that $f_{\rm imcb}$ should be rather high (likely $f_{\rm imcb}>0.8$)
even in GC formation.
We assume that IMCB fraction in GC formation is 1 to avoid introducing
an extra free parameter in the present model,
because the results do not depend strongly on $f_{\rm imcb}$ for the reasonable
range ($0.7<f_{\rm imcb}<1$).

\begin{figure}
\psfig{file=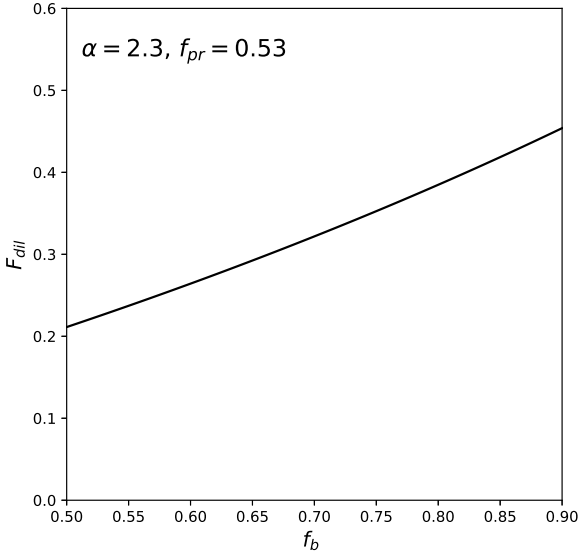,width=8.5cm}
\caption{
Dilution factors ($F_{\rm dil}$) as a function of $f_{\rm b}$
for the models with $\alpha=2.3$ and $f_{\rm pr}=0.53$.
Clearly,  there is almost a linear relation between
$F_{\rm dil}$ and $f_{\rm b}$, and this relation does not depend
so strongly on other parameters such as $\alpha$, $f_{\rm ej}$,
and $f_{\rm pr}$.
}
\label{Figure. 3}
\end{figure}

\begin{figure*}
\psfig{file=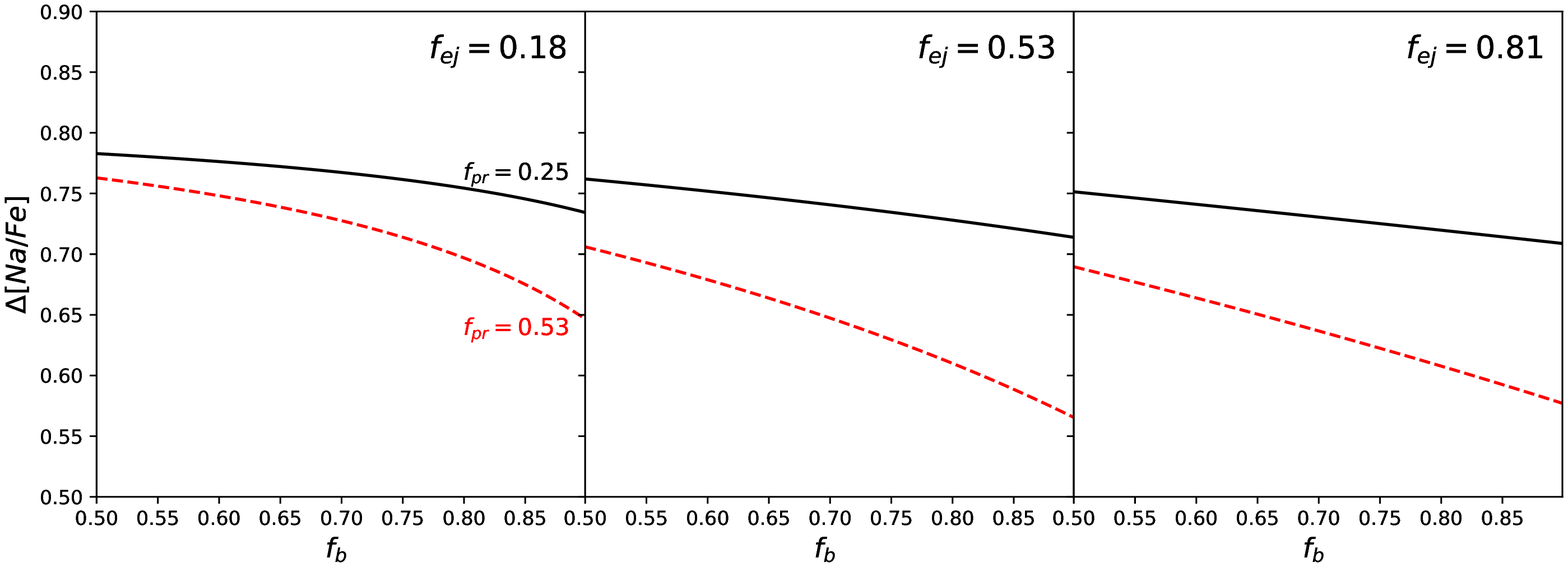,width=18.0cm}
\caption{
Differences in [Na/Fe] between 1G and 2G stars ($\Delta$[Na/Fe]) as a
function of $f_{\rm b}$ in the models with $f_{\rm ej}=0.18$ (left),
0.53 (middle), and 0.81 (right) for $f_{\rm pr}=0.25$ (black solid)
and 0.53 (red dashed).
}
\label{Figure. 4}
\end{figure*}

The IMFs of the single and binary (1G) populations are assumed
to be the same, however, the normalization factor ($C_0$) needs
to be estimated separately for the two using the following
relations:
\begin{equation}
(1-f_{\rm b})M_{\rm i}=\int_{0.25}^{50} C_{0, s} m
\psi (m) dm,
\end{equation}
for the single population, and 
\begin{equation}
f_{\rm b} M_{\rm i}=\int_{0.25}^{ m_{\rm u, 2g} } C_{0, b} (1+q(m)) m
\psi (m) dm,
\end{equation}
for the binary one. In these equations,
the mass-ratio of the primary to the secondary stars in a binary pair
is denoted as $q$ and the  $q(m)$ describes how $q$ depends on
stellar masses.
We use the flat $q$ distribution (V12) and consider no initial
dependence of $q$ on stellar masses. Accordingly,
we adopt $q=0.5$ in the present study.
Thus, once $M_{\rm i}$ and $f_{\rm b}$ are given,
the IMF normalization factors, $C_{\rm 0, s}$ and $C_{\rm 0, b}$
(for single and binary populations, respectively) can be determined
for a given IMF slope $\alpha$.
We do not consider singe and binary populations in 2G stars just
for simplicity in the present study.

The total  mass of gas ($M_{\rm g}$) that can be used
for the formation of 2G stars is the sum of 
(i) the total mass of stellar winds from single AGB stars ($M_{\rm g, s}$)
and (ii) that of gaseous ejecta from IMCBs ($M_{\rm g, b}$).
The IMCB ejecta can be further divided into
(i) pristine gas that can be used to dilute the ejecta
of single AGB stars and (ii) polluted gas from EAGB and TPAGB phases
of IMCBs. 
Following V12, we here consider that the mass fraction of pristine gas
from IMCBs ($f_{\rm pr}$) 
 is the key for the chemical abundance patterns of 2G stars,
and $f_{\rm pr}$ is as follows:
\begin{equation}
f_{\rm pr}=\frac{ M_{\rm g, pr} }{ M_{\rm g, b} }.
\end{equation}
Using the above equations,  we can estimate $M_{\rm g, b}$
for IMCBs with $3 \le m_{\rm s}/{\rm M}_{\odot}  \le 8$
as follows:
\begin{equation}
M_{\rm g, b} =f_{\rm ej} \int_{3}^{8} C_{0, b} (1+q(m)) m
\psi (m) dm,
\end{equation}
where $f_{\rm ej}$ is the average mass fraction of gas ejected
from IMCBs (V12). Since both $f_{\rm pr}$ and $f_{\rm ej}$ are
calculated in V12 for a large number of models,
we can use the derived values to estimate $M_{\rm g, pr}$ in the 
above equations.

In order to calculate $M_{\rm g, s}$, 
we use the following analytic form for the total mass of stellar wind 
($m_{\rm w}$)
from a single AGB star (G19):
\begin{equation}
m_{\rm w} =0.894 m-0.434,
\end{equation}
which is only slightly different from our previous works (Bekki 2011, B11).
Accordingly, $M_{\rm g, s}$ can be estimated as follows:
\begin{equation}
M_{\rm g, s}=\int_{3}^{8} C_{\rm 0, s} m_{\rm w}
\psi (m) dm.
\end{equation}
The sum of this $M_{\rm g, s}$ and $(1-f_{\rm pr})M_{\rm g, b}$ is
the total mass of ``polluted'  gas.
The mass fraction of pristine gas among all gas is referred
to as a ``dilution factor'' ($F_{\rm dil}$),
which is defined as follows:
\begin{equation}
F_{\rm dil}=\frac{ f_{\rm pr} M_{\rm g, b} }{ M_{\rm g} }.
\end{equation}

We investigate (i) possible differences in [Na/Fe] between
1G and 2G stars and (ii) possible Li abundance (A(Li)) of 2G stars
in this SBC scenario in order to address the validity of the scenario.
In order to discuss these two problems more quantitatively,
we simply adopt the following assumptions.
First, [Na/Fe] for 1G (${\rm [Na/Fe]}_{\rm min}$)  and polluted gas from AGB stars
(${\rm [Na/Fe]}_{\rm max}$)  
are $-0.1$ and 0.7, respectively.
These are adopted as typical values of GCs with multiple stellar populations
and consistent with observations, e.g., the results shown in Fig. 1 by
Carretta et al. 2010.
Second,  pristine gas from IMCBs has no Li (``Li-free gas''), though this could lead to
an underestimation of A(Li) of 2G stars.
Here we follow the models by D'Antona et al. (2012, D12), who assumed
Li-free gas in some of their models and thereby investigated the possible
A(Li) of 2G stars in GCs.
Third, the same Li yields ($Y_{\rm Li}$) adopted by D12 are used in the present study
so that the mean A(Li) of AGB ejecta can be calculated.
The mean mass-weighted  fraction of Li among AGB stars
($Y_{\rm Li, m}$)  is estimated as follows:
\begin{equation}
Y_{\rm Li, m}=
\frac{1}{ M_{\rm g, s} }
\int_{3}^{8} C_{\rm 0, s} Y_{\rm Li} (m) m_{\rm w}
\psi (m) dm 
\end{equation}
where $Y_{\rm Li}(m)$ is dependent on the masses of AGB stars
and adopted from D12.
Based on this $Y_{\rm Li, m}$, we estimate the mean A(Li) for AGB ejecta
($A_{\rm m}(\rm Li)$). The value of  ${\rm A}_{\rm m}({\rm Li})$ is 2.4 for $\alpha=2.3$, and it
does not depend on $\alpha$.

The adopted assumption of Li-free pristine gas from IMCBs 
would be  oversimplified,
because Li fraction is lower yet not zero 
in pre-AGB phases of intermediate-mass stars (see Fig. 9 in 
Karakas and  Lattanzio 2014). The observed A(Li) in stars of  NGC 6397 and
predicted one 
range from $\approx 2$
to $\approx 1$
even during first dredge-up phases  of intermediate-mass stars
(Karakas and  Lattanzio 2014). Therefore, the  present model 
is highly likely to under-predict A(Li) of 2G stars.

\begin{figure*}
\psfig{file=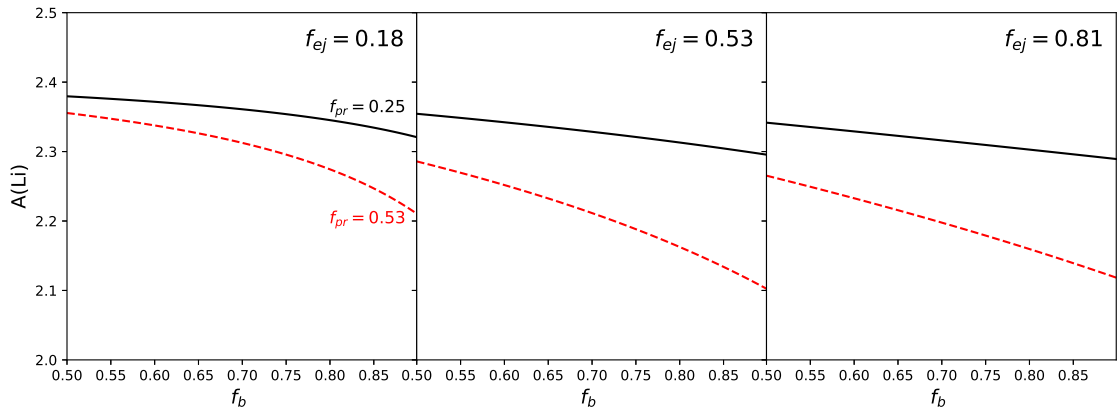,width=18.0cm}
\caption{
The same as Fig. 4 but for A(Li) of 2G stars. 
${\rm A}_{\rm m} {\rm (Li)}=2.4$ predicted from D12 is adopted
in these models.
}
\label{Figure. 5}
\end{figure*}

\subsection{$f_{\rm 2g}$}

Using the above simple analytic models, we can discuss
how $f_{\rm 2g}$ depends on the model parameters. Fig. 1 demonstrates that
there is a narrow range of $\alpha$ for which
$f_{\rm 2g}$ can be larger than the observed $f_{\rm 2g}$
($\approx 0.4$)  
in the Galactic GCs with $\ log(M_{\rm gc}/{\rm M}_{\odot})=5.2$
(G19). 
For example, the required $\alpha$ for $f_{\rm 2g} > 0.4$
in the models with $m_{\rm u, 2g}=1 {\rm M}_{\odot}$  is $\approx 1.8$
for $f_{\rm b}=0.5$ and $\approx 1.7$ for $f_{\rm b}=0.7$, which means that
IMFs of 1G stars need to be moderately top-heavy.
There is no $\alpha$ that can reproduce $f_{\rm 2g}=0.4$ in the models
with $m_{\rm u, 2g}=8 {\rm M}_{\odot}$,
which implies that high-mass star formation needs to be severely suppressed in
the 2G formation.

Clearly, $f_{\rm 2g}$ is larger for smaller $\alpha$ (i.e., more top-heavy IMFs)
for different $f_{\rm b}$, mainly because the mass fractions of single AGB
and IMCB stars are larger for smaller $\alpha$. Slightly larger $f_{\rm 2g}$
in smaller $f_{\rm b}$ is due to the adopted assumption that the mass fractions
of gaseous ejecta  among single AGB stars (more than 0.8)
are larger than those of gas among
IMCBs ($f_{\rm ej}=0.53$). 
In these models with $\alpha=2.3$ (standard Salpeter IMF),
$f_{\rm 2g}$ cannot be larger than 0.4 for any $m_{\rm u, 2g}$ and $f_{\rm b}$.
This suggests that 1G stars with $\alpha=2.3$
 need to be much more efficiently lost compared with
2G through some physical processes (e.g., tidal stripping by their host dwarf
galaxies) to reproduce the observed typical $f_{\rm 2g}$. Such preferential
stripping of 1G stars has been already proposed in D08 and investigated
by Vesperini et al. (2010) for the standard AGB scenario.

The required $\alpha \approx 1.8$ is not unrealistic, given that
the IMF slopes for intermediate- and high-mass
stars are inferred to be  top-heavy ($\alpha<2.0$) in
70\% of the investigated 20 GCs (Marks et al. 2012, M12).
However, it should be noted here that
the inferred IMF slopes in M12 are based on the assumed link between
the IMF slope and the gas expulsion process.
The various physics in young star clusters can be possibly missed
in their inference of IMF slopes from observation: see Krause et al. (2020)
for a more recent
review of various physical processes in young star clusters.

Fig. 1 also indicates that the observed dispersion in $f_{\rm 2g}$
can be due to the dispersion in $\alpha$ between GCs: M12 indeed
showed a dispersion in $\alpha$.
Necessity  of top-heavy IMFs to explain the observed large
$f_{\rm 2g}$ was discussed in other scenarios of GC formation (e.g.,
Bekki \& Norris 2006;
Prantzos \& Charbonnel 2006), 
which implies that
$f_{\rm 2g}$ can provide a constraint on the IMF in GC formation.

\subsection{$F_{\rm mb}$}

Fig. 2 shows that $F_{\rm mb}$ (``mass budget factor'')
ranges from $\approx 2$ at $\alpha=2.5$ 
to $\approx 7$ at $\alpha=1.5$ for $m_{\rm u, 2g}=8 {\rm M}_{\odot}$.
As expected, $F_{\rm mb}$ is larger for smaller $\alpha$
(i.e., more top-heavy IMFs), because a larger amount
of gas from CCSNe is lost and a large number of massive stars are locked as
compact objects (stellar mass black holes and neutron stars) for smaller
$\alpha$.
This $\alpha$-dependence can be clearly seen in 
Fig. 2 for the models with different $m_{\rm u, 2g}$,
and it is confirmed to be seen in the models with different $f_{\rm b}$,
though the results are not shown in Fig. 2.
Based on the detailed comparison of GC properties between new Gaia DR2 data
and corresponding simulations,
Baumgardt et al. (2019) showed that GCs have lost about 80\% of its initial
mass on average (i.e., $F_{\rm mb} \approx 5$).
The derived maximum possible $F_{\rm mb}$ ($\approx 7$) in
the present study is therefore
not too large (like $10-100$
suggested in other previous works),
which implies that there would not be 
a  serious ``mass-budget'' problem in this SBC scenario.
Even for $\alpha<1.7$ required for $f_{\rm 2g}>0.4$  in Fig. 1, 
$F_{\rm mb}$ is $\approx 4$  for $m_{\rm u, 2g}=1 {\rm M}_{\odot}$.

It should be noted here that all gas ejected from single AGB stars and IMCBs 
are assumed to be converted into 2G stars in these models. 
Our recent hydrodynamical simulations of 2G formation in dense stellar systems
have shown that
such a 100\% gas conversion efficiency is unlikely (B11, 
Bekki 2019, B19b). Accordingly,
$F_{\rm mb}$ in Fig. 2  can be the lower limit for each $\alpha$.
In this discussion,   loss of stars due to long-term internal dynamical
relaxation processes and tidal stripping of stars by GC host dwarfs
are not considered at all. Therefore, the initial masses of GCs can be even
larger than $M_{\rm i}$ estimated in the present study.

\subsection{$\Delta$[Na/Fe]}

Fig. 3 describes how $F_{\rm dil}$ (dilution factor)
depends on $f_{\rm b}$ for a fixed $\alpha$ and $f_{\rm pr}$.
A larger amount of pristine gas can be ejected from IMCBs for larger $f_{\rm b}$
so that gaseous ejecta from single AGB stars can be diluted to a higher degree
(i.e., larger $F_{\rm dil}$). This result does not depend 
on $\alpha$ and $m_{\rm u, 2g}$, but it depends strongly on $f_{\rm pr}$ and
$f_{\rm ej}$ for IMCBs. 
The derived $F_{\rm dil}$-dependence suggests that
abundance differences between 1G and 2G stars can be larger for GCs with 
smaller $f_{\rm b}$.
Indeed, as demonstrated in Fig. 4, 
$\Delta$[Na/Fe] can be larger for smaller $f_{\rm b}$ for different $f_{\rm ej}$
and $f_{\rm pr}$, though the $f_{\rm b}$ dependence is not so strong.
Clearly, $\Delta$[Na/Fe] is systematically
lower in the models with larger $f_{\rm pr}$ 
in which a larger amount of pristine gas can be ejected from IMCBs for a given
$f_{\rm ej}$.

The key parameter here is 
$f_{\rm b}$ at the epoch  when intermediate-mass stars enter into
their AGB phases.  As shown in previous numerical simulations
(e.g., Hong et al. 2015),
$f_{\rm b}$ in GCs can be dramatically reduced due to 
the internal  dynamical relaxation processes.
If the rapidity of $f_{\rm b}$ reduction can be simply scaled to
the dynamical relaxation
timescale at half-mass radii of  GCs ($t_{\rm relax}$),
then $f_{\rm b}$ at AGB phases of intermediate-mass stars
can be lower for GCs with shorter $t_{\rm relax}$.
This suggests that there could be an anti-correlation
between $t_{\rm relax}$ and $\Delta$[Na/Fe] in GCs.
This possible anti-correlation could disappear, if
$f_{\rm pr}$ and $f_{\rm ej}$ are different between GCs
with different $t_{\rm relax}$ for some physical reasons.

\begin{figure*}
\psfig{file=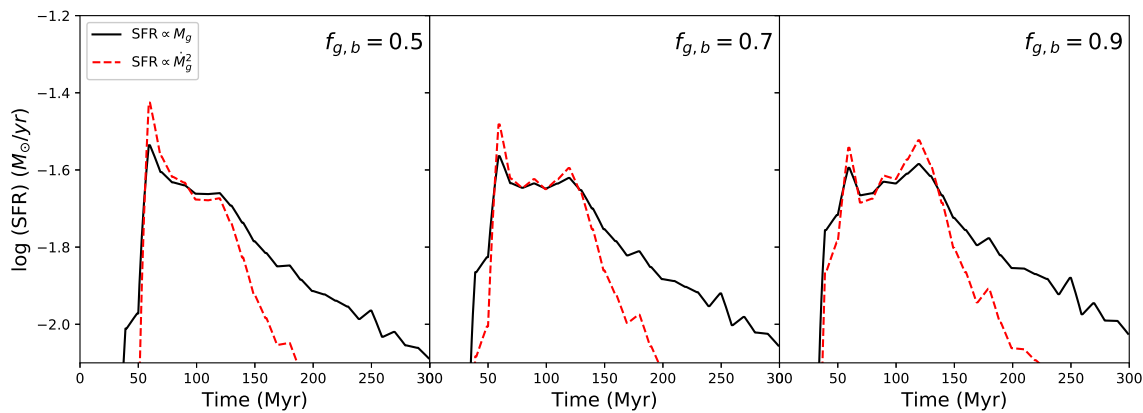,width=18.0cm}
\caption{
Time evolution of SFRs in the models with $f_{\rm g, b}=0.5$ (left),
0.7 (middle), and 0.9 (right) for mass-dependent (i.e.,
SFR $\propto  M_{\rm g}$; black solid)
and accretion-dependent star formation laws 
(i.e., SFR $\propto$ {\it \. M}$^2$;
red dashed).Two peaks can be more clearly seen in the model 
with $f_{\rm g, b}=0.7$ and 0.9 for which a larger amount of pristine gas
can be accumulated in GCs.
The two peaks are due to the two peaks in the mass ejection rates of
IMCBs (V12). These results imply that 2G stars can have two distinct
populations for larger  $f_{\rm g, b}$.
}
\label{Figure. 6}
\end{figure*}

\subsection{A(Li) of 2G stars}

As shown in Fig. 5, the predicted dependence of A(Li) on $f_{\rm b}$ is very
similar to $f_{\rm b}$-dependence of  $\Delta$[Na/Fe]. The models
with larger $f_{\rm b}$  have lower A(Li) in their 2G stars owing to
larger degrees of dilution of AGB ejecta by pristine gas from IMCBs.
The $f_{\rm b}$-dependence is rather flat for smaller $f_{\rm ej}$ and $f_{\rm pr}$,
because only a small amount of pristine gas can be ejected from IMCBs and subsequently
mixed with AGB ejecta. 
There is not a significant difference in A(Li) of 2G stars between
models with different $f_{\rm b}$ and $f_{\rm ej}$ for $f_{\rm pr}=0.25$.
In these
models, A(Li) of 2G stars can be determined largely by the Li yields of AGB
stars  (${\rm A}_{\rm m}{\rm (Li)} = 2.4$).
Irrespectively of $f_{\rm b}$ and $f_{\rm ej}$, 
the models with large $f_{\rm pr}$ show lower A(Li).

In each model, $f_{\rm pr}$ averaged over a IMCB population from V12
is used. However, $f_{\rm pr}$ should be time-dependent in real GC formation,
because IMCBs with different $m$, separations, and orbital periods can have
different $f_{\rm pr}$ (V12). Since pristine gas can be ejected from IMCBs  earlier
than polluted one, it is possible that intracluster gas can have rather low
A(Li) in the very early formation phases of GCs, e.g., $\approx 30$ Myr after
GC formation, i.e., when
single AGB stars with $m=8 {\rm M}_{\odot}$ start to eject polluted gas.
Accordingly, 2G stars formed earlier can have lower A(Li) and
[Na/Fe] similar to that of 1G stars. Furthermore,
the fractions of Li-poor stars can be larger in GCs in which 2G star formation
can start earlier.
One of possible predictions in the SBC scenario is that there can be 2G stars with
rather low A(Li) formed from almost Li-free gas.
Such formation of Li-poor 2G stars can be suppressed if there is a threshold
gas mass or density beyond which star formation is possible,
because intracluster 
gas masses/densities can significantly increase as gas from single AGB stars
is accumulated within GCs.

\section{Star formation histories of 2G stars}

In this section, we use classic one-zone models to
investigate the possible star formation histories (SFHs) of 2G stars in
forming GCs. Although we used hydrodynamical  simulations in our
previous works on star formation rates (SFRs) within
GCs (B11; Bekki 2017, B17; B19b),  we consider that the present one-zone models
are enough to provide useful information on the possible SFHs 
in this first paper.
Such more sophisticated
(and time-consuming) modeling will  be done in our forthcoming
papers to discuss several other key issues that are not discussed in the present
paper.
It should be stressed here that previous observational and theoretical studies
of the formation of individual stars focused exclusively on star formation
within molecular clouds. Accordingly, it is totally unknown how star formation
can proceed in  molecular gas  embedded in dense stellar systems. Therefore,
the present-study makes the most of the models used in previous studies of
galactic star formation and in numerical simulations
of globular cluster formation (B19b).

\subsection{One-zone models}
We investigate the time evolution of $M_{\rm gc, g}$ (total gas mass)
and SFR ($\psi(t)$) in a GC
for a given
accretion rate ($A(t)$)  of gas from single and binary stars
within GCs.
The basic equations for the adopted one-zone  models of SFHs in 2G stars
are described as follows:
\begin{equation}
\frac{d M_{\rm gc, g }}{dt}=-\psi(t)+A(t).
\end{equation}
Here we distinguish between $M_{\rm gc, g}$ and $M_{\rm g}$, because
$M_{\rm gc, g}$ is the sum of gas from all 1G stars and ISM from GC host galaxies
in GC formation:
ISM accretion onto GCs is not considered in the present study, however.
Since this gas accretion is both from AGB winds of single stars and from
mass loss of IMCBs due to RLOF, $A(t)$ can be defined as follows:
\begin{equation}
A(t)=
\frac{d M_{\rm g }}{dt}=
\frac{d M_{\rm g, s }}{dt}+
\frac{d M_{\rm g, b }}{dt}.
\end{equation}

We can calculate 
$dM_{\rm g, s }/dt$ using the adopted IMF as follows:
\begin{equation}
\frac{d M_{\rm g, s }}{dt}
=\frac{1}{dt} \int_{ m_{\rm to} (t+dt) }^{ m_{\rm to} (t) } C_{\rm 0, s} m_{\rm w} 
\psi (m) dm,
\end{equation}
where $m_{\rm to}$ is the main sequence turn-off mass, which is a function
of ages of stars (i.e.,  time $t$) and always
$m_{\rm to}(t) > m_{\rm to} (t+dt)$.
In order to calculate $m_{\rm to}$ at each time $t$, we use the following
relation between stellar  ages ($t_{\rm s}$) and $m_{\rm to}$ by 
(Greggio  \& Renzini 2011):
\begin{equation}
\log m_{\rm to}(t_{\rm s})
= 0.0434 (\log t_{\rm s})^2 - 1.146 \log t_{\rm s} + 7.119,
\end{equation}
where $m_{\rm to}$  is in solar units and time $t_{\rm s}$ in years.

We use the results of the models shown in V12 (their Fig. 2) to estimate
$dM_{\rm g, b}/dt$ for all models. Using the table kindly provided D. Vanbeveren,
we first estimate $dM_{\rm g, b}/dt$ for each time step with $dt =10^6$ yr
from the table and then normalize it for a given initial mass of
a GC. A key parameter here is the mass fraction ($f_{\rm g, b}$) 
of gas ejected from IMCBs over $300$ Myr and  among all gas ejected
from 1G stars over the same period, which is defined as follows;
\begin{equation}
f_{\rm g, b}=\frac{ M_{\rm g, b} }{ M_{\rm g} }.
\end{equation}
Here it should be noted that this is not simply a fraction of binary stars,
but it depends both on the adopted IMFs and on
a number of parameters that describe
the time evolution of IMCBs (V12). Although we have  investigated
a number of different $f_{\rm b, g}$, we mainly show the results
of models with $f_{\rm g,b}=0.5$, 0.7, and 0.9,
because the adopted $f_{\rm g, b}$ values
are quite reasonable.

We adopt two star formation models, because it is not clear how
individual stars can be formed within such dense environments of 1G stars.
One model assumes  that
the star formation rate $\psi(t)$ is proportional
to $M_{\rm g}^{\beta}$ (``mass-dependent''), where $\beta$ is   
a  power-law slope,
and a constant star formation
coefficient ($\epsilon_{\rm sf}$)  is adopted.
Thus it is described as follows:
\begin{equation}
\psi(t)=C_{\rm sf}M_{\rm g}^{\beta}(t)
\end{equation}
where $C_{\rm sf}$ is a constant that determines SFRs.
This constant is determined such that most of the accreted gas
can be converted into 2G stars for each model.
This star formation model dependent on gas mass or gas mass fraction (density)
has been adopted often in the studies of galactic chemical evolution
using so-called ``one-zone models''.
However, it is not clear at this stage whether such star formation models 
can be applied for 2G formation in forming GCs.
We thus suggest that the real star formation histories of 2G could be 
significantly different from those derived using the above model.

The other assumes that gas accretion rates  can determine SFRs
as follows:
\begin{equation}
\psi(t)=C_{\rm sf}A(t)^{\beta}
=C_{\rm sf}{(dM_{\rm g}/dt)}^{\beta} .
\end{equation}
In addition to the mass-dependent model, this model is also investigated,
because we consider that gas accretion rates can be important for the 
formation of high-density
cores where star formation can occur. Again, this model is just one of possible 
star formation models, and accordingly, the real secondary star formation
can be different from the predicted one from the model.

We here newly introduce a hypothetical ``threshold gas mass'' ($M_{\rm g, th}$)
beyond which star formation is possible. This introduction of $M_{\rm th, g}$
is motivated by our previous simulations (B19b), which shows that star formation
within GCs can be severely suppressed in lower gas mass fractions
(i.e., lower $M_{\rm gc, g}/M_{\rm gc}$) owing to the direct collisions of
cluster member stars with gas clouds.
This suppression mechanism of star formation due to interaction between
stars and gas clouds is very unique in dense stellar systems (not relevant
to star formation within molecular clouds without existing stars).
Accordingly  it is better for the present study to introduce $M_{\rm g, th}$ at least
in some models.
In the model with the threshold gas mass,  
\begin{equation}
\psi(t)=0
\end{equation}
for $M_{\rm gc, g} < M_{\rm g, th}$.
Once $M_{\rm gc, g} \ge M_{\rm g, th}$, 
gas starts to be consumed following the adopted star formation model described
above.
We here consider that (i) star formation efficiency (SFE, $\epsilon_{\rm sf}$)
is fixed 
and (ii) this star formation can continue 
until a gas mass of  $\epsilon_{\rm sf} M_{\rm g, th}$ (which corresponds
to $\epsilon_{\rm sf}M_{\rm gc, g}$ when $M_{\rm gc, g}$
becomes $M_{\rm g, th}$)
is all converted into
new stars.
We assume a high SFE of $\epsilon_{\rm sf}=0.7$ in the present models.
Accordingly, if $M_{\rm g, th}=10^5 {\rm M}_{\odot}$,
then gas can be consumed until $M_{\rm gc, g}$ becomes 
$3 \times 10^4 {\rm M}_{\odot}$.
In the mass-dependent models, star formation needs to be truncated by
some physical processes: otherwise, very low-level star formation can produce
2G stars with their chemical abundances influenced by low-mass AGB stars.
Feedback effects of SNIa that can occur frequently within 0.1-1 Gyr after
1G formation can remove all of the remaining gas from forming GCs.
However, without detailed hydrodynamical simulations of feedback effects from 
SNIa, it would not be possible to claim that SNIa can cause such
truncation of star formation.

\begin{figure}
\psfig{file=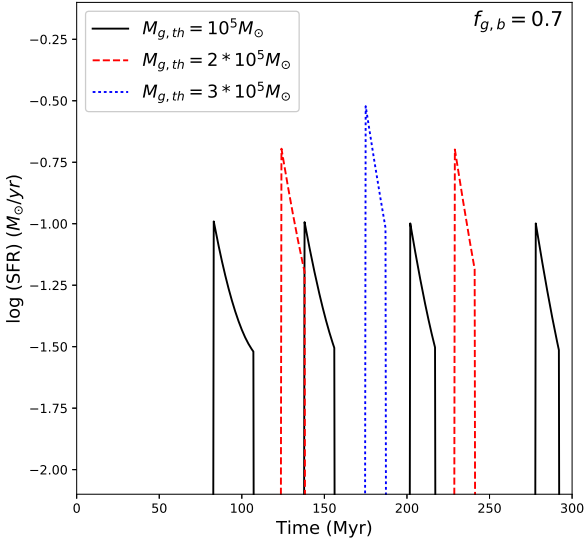,width=8.5cm}
\caption{
Time evolution of SFRs in the models 
with $M_{\rm g, th}=10^5 {\rm M}_{\odot}$ (black solid),
$2 \times 10^5 {\rm M}_{\odot}$ (red dashed),
and $3 \times 10^5 {\rm M}_{\odot}$ (blue dotted).
The adopted $M_{\rm g, th}$ are adopted so that the formation
of 2G, 3G, 4G etc due to $M_{\rm g, th}$ can be demonstrated.
}
\label{Figure. 7}
\end{figure}

\subsection{Results}

Fig. 6 describes the time evolution of SFRs over 300 Myr
in the models with different
$f_{\rm g, b}$ in which the total gas mass ejected from single AGB stars
and IMCBs is $6 \times 10^5 {\rm M}_{\odot}$. 
In these models, SFRs at each time step are calculated
using the mass-dependent (SFR $\propto M_{\rm g}$ with $\beta=1$) 
and accretion-dependent SF models (SFR $\propto$ {\it \. M}$^2$ 
with $\beta=2$)
for the adopted V12 model  for the time evolution rate of gas from
IMCBs. Clearly, there are two peaks in SFR evolution for models with
$f_{\rm g, b}=0.7$ and 0.9, which reflects the fact that there are two peaks
in the time evolution of gas ejection rates for IMCBs.
The two peaks are more distinct for the accretion-dependent star formation
model, simply because  SFR is more sensitive
to $dM_{\rm g}/dt$ that is determined largely
by gas ejection rates of IMCBs in the model.
The two SFR peaks can end up with two stellar populations
with different chemical abundances in 2G stars owing to the possible
difference in chemical abundance patterns of intracluster gas at the two epochs.
We will investigate this point in the context of discrete MPs 
in our future works.

Star formation can be naturally almost truncated around 200 Myr
after 1G formation in the accretion-dependent
models, whereas it can only slowly decline 100 Myr after 1G formation in
the mass-dependent model. As pointed out for the ``truncation'' dilution
problem in the standard AGB scenario, gas from AGB stars with $m<3{\rm M}_{\odot}$ 
should not contribute to chemical enrichment of intracluster gas from which
2G stars are formed, because C+N+O of 2G stars can be significantly different
from 1G. Therefore the SBC scenario with the mass-dependent SF has a
similar truncation dilution problem.
Fig. 6 demonstrates that SFRs are rather low over the 2G formation
($\log {\rm SFR} < -1.4 {\rm M}_{\odot}$ yr$^{-1}$). If the 
relation between maximum stellar masses ($m_{\rm max}$)  and SFRs
used in Bekki et al. (2017) is adopted for these SFRs,
then $m_{\rm max}$ (i.e., $m_{\rm u, 2g}$) can be significantly lower
than $50 {\rm M}_{\odot}$ (adopted for 1G), 
but not low enough to explain the origin of the required $m_{\rm u, 2g}$
($=8 {\rm M}_{\odot}$) in the SBC scenario.

Fig. 7 demonstrates that multiple sharp peaks in SFR evolution
with wide SFR intervals  can be reproduced in 
the mass-dependent star formation models with  $f_{\rm g, b}=0.7$ and
$M_{\rm g, th} \ge 10^5 {\rm M}_{\odot}$.
These values of $M_{\rm g, th}$ are adopted so that the intervals
between multiple episodes of star formation can be more clearly
seen: there is little physical basis for the adopted values.
The mass-ratio of $M_{\rm g, th}$ to $M_{\rm i}$ rather than
simple $M_{\rm g, th}$ would be a more physically meaningful parameter,
because self-gravity can play a role in star formation
in dense stellar systems (B19b).

In these models, the number of peaks in SFR evolution 
depends on $M_{\rm g, th}$ such that it can be larger for smaller $M_{\rm g, th}$.
For example, the formation of 2G, 3G, 4G and 5G stars is possible
for $M_{\rm g, th}=10^5 {\rm M}_{\odot}$ whereas
only 2G formation is possible for
$M_{\rm g, th}=5 \times 10^5 {\rm M}_{\odot}$.
Although these results imply that the origin of discrete MPs observed in
some GCs can be understood in the context of $M_{\rm g, th}$ for 2G formation,
it is theoretically unclear which value of $M_{\rm g, th}$ is 
the most reasonable and realistic.

In order for 2G stars with $f_{\rm 2g} \approx 0.4$
to be formed in the SBC scenario,
a long duration (at least an order of $10^8$ yr) of star formation is required
in these models.
This would be possible only if both delayed CCSNe formed from IMCBs (V12)
and SNIa do not occur at all 
during 2G formation (or at least their formation is severely suppressed).
It would be equally possible that 2G formation from gas ejected from single 
AGB stars and IMCBs can occur only after all delayed CCSNe from IMCBs are exploded.
In this case, the total mass of 2G stars can be significantly reduced because
gas ejected earlier from massive AGB stars and IMCBs 
can be expelled by delayed CCSNe completely from GCs: $f_{\rm 2g}$ can become
even lower.
As suggested by Krause et al. (2020), 
a significant
number of prompt SNIa can cause steady winds and thereby remove intra-cluster gas in
massive star clusters with masses more than $10^6 {\rm M}_{\odot}$
only 100 Myr after 1G formation.
Accordingly, $f_{\rm 2g}$ should also depend strongly on the delay time distribution
of SNIa in forming GCs.

\section{Discussion}

\subsection{Origin of $f_{\rm 2g}$ dependent on $M_{\rm gc}$}

The observed mass fractions of 2G stars ($f_{\rm 2g}$) in GCs are larger 
in more massive GCs (G19, MM22). Although the origin of this $f_{\rm 2g}-M_{\rm gc}$
relation is  yet to be fully understood,
MB21 have demonstrated that this relation can be reproduced well
in the simulated GCs, 
if accretion of AGB ejecta and ISM is included in their GC formation models.
Such results are reasonable, because accretion of ISM onto GCs depends
on $M_{\rm gc}^2$ for the Bondi-type accretion for a given set of
model parameters of ISM in GC host galaxies, and secondly because
the mass of AGB ejecta depends on $M_{\rm gc}$ for a given IMF.
However, it would be important and instructive for the present study
to discuss whether the  SBC scenario (without
ISM accretion onto GCs) can provide an explanation for
the $f_{\rm 2g}-M_{\rm gc}$ relation too.

As shown in the present study, $f_{\rm 2g}$ depends strongly on $\alpha$
in such a way that $f_{\rm 2g}$ is larger for more top-heavy IMFs
(i.e., flatter IMFs). 
This is because the mass fraction of AGB stars (thus the fraction
of AGB ejecta) is larger for
flatter IMFs whereas the mass fraction of low-mass 1G stars
is smaller for such IMFs (thus the relative fraction of 2G low-mass
stars is larger).
This implies that if more massive GCs have
flatter IMFs during their 1G formation, then the observed
$f_{\rm 2g}-M_{\rm gc}$ can be qualitatively reproduced at least. 
Young  GCs with multiple stellar populations
at high $z$ are yet to be discovered.
Even if the possible GC candidates in high-$z$ dwarfs are discovered,
it would be  currently almost impossible for observational studies to 
investigate the IMFs of 1G stars for such unresolved high-$z$ GCs.
However, the IMFs inferred from mass functions of low-mass
stars for the present-day GCs (e.g., De Marchi et al. 2007;
M12) can be used for this discussion.

In particular,  M12 showed that the IMF slope ($\alpha_3$) relevant to
the formation of intermediate-mass and massive stars can be flatter for 
more massive GMCs from which GC were formed.
Although their results are not the direct observational evidence,
they imply  that if SFEs are not
so different between different GMCs, then more massive GCs are likely to
have flatter IMFs: indeed, some of more massive GCs (e.g., NGC 5139) in their
Table 1 shows rather flat IMF slopes ($\alpha_3 < 1.7$).
Although we can propose  that the $\alpha-M_{\rm gc}$ (or more precisely,
$\alpha-M_{\rm i}$) relation can be responsible for
the observed $f_{\rm 2g}-M_{\rm gc}$ relation,
this proposal needs to be investigated more quantitatively
using next-generation sophisticated 
numerical simulations of GC formation in our future papers.

\subsection{The mass budget problem}

The mass budget factor ($F_{\rm mb}$) in the SBC scenario ranges
from 2 to 7, depending on $\alpha$ and $m_{\rm u, 2g}$. 
As shown in Fig. 2,  if $\alpha$ is larger than 1.8, then $F_{\rm mb}$
can be smaller than $\approx  5$ even for $m_{\rm u, 2g}=8 {\rm M}_{\odot}$. 
Recent $N$-body simulations of long-term GC evolution by Webb and Leigh (2015)
have demonstrated that the initial masses of GCs are typically 4.5 times
larger than their  present-day masses for the adopted Kroupa IMF.
Although these simulations include long-term dynamical effects on
the mass loss processes of GCs, the present study did not include
such effects
in estimating $F_{\rm mb}$:
these two studies cannot be simply compared each other.
However, the results by Webb and Leigh (2015)
strongly suggest that derived $F_{\rm mb}$ in the SBC scenario
is quite reasonable and realistic.
Accordingly, there is no serious ``mass-budget'' problem in the SBC scenario.
However, it should be noted here that all gas from single AGB and
IMCBs is assumed to be consumed by secondary star formation 
for the estimation of $F_{\rm mb}$ in the present study.
Therefore, the  derived $F_{\rm mb}$ could be underestimated.

In discussing the mass budget problem, 2G stars were assumed to form from
ejecta of 1G stars in previous works.
However,  this assumption could  be oversimplified for the following
reasons. 
First, B19a has shown that about 40\% of 2G stars in GCs
are formed from ejecta of  AGB stars that are not 1G stars 
of the GCs but are from field stars formed around the GCs at the
same formation epochs of  the 1G stars.
Second,
observations showed that star clusters can form as cluster associations
(e.g., Bastian et al 2005), which implies that GCs were also formed
with other surrounding  smaller star clusters.
It would be possible that gas ejected from evolved
stars in  the smaller clusters (that do not finally
become the 1G stars of the GCs but become field stars)  could be
trapped by the GCs to be converted into 2G stars.
Therefore, gas from 1G stars of GCs is not the only source 
for secondary star formation: the oversimplified assumption
adopted in previous studies 
made $F_{\rm mb}$ quite large ($>10$).
Thus, it is possible that the mass budget problem is not so serious
as ever thought in previous works (e.g., BL18).

\subsection{Discrete multiple stellar populations}

Recent observations have revealed  that some of the Galactic
GCs have clearly distinct distributions of stars in the [Na/Fe]-[O/Fe]
and [Al/Fe]-[Mg/Fe] diagrams (e.g., Carretta et al. 2012;  Johnson et al. 2019).
Although a few theoretical models to reproduce
the observed discreteness of MP have been 
proposed so far  (e.g., Bekki et al. 2017;
Kim \& Lee 2018; Johnson et al. 2019), 
its origin is yet to be fully understood.
The present study has demonstrated that there can be 
two peaks in SFHs of 2G stars, which
implies that 2G stars can have two distinct major populations.
Furthermore,
the present study has shown that if there is a threshold gas mass ($M_{\rm g, th}$)
for star formation,  GCs can have multiple discrete epochs of 2G formation.
Since AGB stars with different mass ranges thus different stellar yields
can enrich the intracluster medium at different epochs of star formation,
it is possible that 2G populations formed at different epochs
can have different chemical abundances.

Previous works (e.g., Bekki et al. 2017; Kim \& Lee 2018)
 are all based on one-zone models, for which model
parameters can be fine-tuned to match the observed properties of GCs.
Therefore, it is not so clear whether the ranges of physical conditions
required for explaining the discreteness of MPs in GCs in these models
can be achieved
in real GC formation.
Our previous simulations of GC formation with a model for a threshold gas
density for star formation shows multiple epochs of star formation 
due to truncation of star formation by CCSNe (B17).
However, the adopted threshold density of $10^4-10^5$ cm$^{-3}$ 
is yet to be justified by further theoretical investigation
of star formation in dense environments of 1G stars.
Furthermore, no numerical simulations have ever investigated whether or not
gas from AGB stars and massive OB stars, which should be formed at two different
epochs (i.e., before and after CCSNe), can really mix well with
gas left over from 1G formation to finally form new
stars in the scenario proposed by Kim \& Lee (2018).

The present one-zone models did not include
the time evolution of chemical abundances of
O, N,  Mg, and Al 
and thus is unable to discuss the origin of the observed distinct populations 
on the [Na/Fe]-[O/Fe]
and [Al/Fe]-[Mg/Fe] diagrams.  
Also so far no numerical simulations of GC formation have clearly demonstrated
the distinct clumpy distribution of the stars on the [Na/Fe]-[O/Fe]
and [Al/Fe]-[Mg/Fe] diagrams.  Thus the main aim 
of our future more sophisticated simulations is 
to reveal (i) the physical origin of the hypothetical
threshold gas mass or density for star formation in dense stellar systems
and  (ii) the roles of the threshold mass/density in reproducing the 
formation of discrete MP in the two diagrams.

\subsection{Avoidance of delayed 1G CCSNe and suppression of 2G massive star formation}

Like the standard AGB scenario (e.g., D08, D12),
the SBC scenario requires severe suppression of 
massive star formation with $m \ge 8 {\rm M}_{\odot}$ leading
to  CCSNe during 2G formation
and SNIa, firstly because
such energetic events can expel most  gas for 2G formation
(e.g.,   Lacchin et al. 2021),
and secondly because they can introduce large [Fe/H] spreads ($>0.3$)
that are not observed in Type I GCs (e.g., G19, MM22).
Accordingly, $m_{\rm u, 2g}$ needs to be lower than $8 {\rm M}_{\odot}$
for at least $\approx 100$ Myr in the SBC scenario.
A key question is therefore whether and how such massive star formation can be almost
completely suppressed in the dense environments of GCs.

Using hydrodynamical simulations of GC formation,
B19b investigated how frequent dynamical interactions between (1G) stars
and cold gas (``stellar bombardment'')  influence the formation of 
gas clouds leading to star formation, and found that
gas clouds more massive than $\approx 3 {\rm M}_{\odot}$ can be completely
truncated by such stellar bombardment effects (e.g., Fig. 11 of B19b).
The simulations, however, do not include various physical effects on
star formation such as the radiation fields of 1G stars, magnetic fields,
dust physics in intracluster gas of GCs. Accordingly, more sophisticated
numerical simulations of 2G formation are required to confirm the
low $m_{\rm u, 2g}$. 

De Donder \& Vanbeveren (2004) showed that 
initially intermediate-mass stars in binaries can be transformed into massive stars
that can explode as CCSNe and therefore that CCSNe can be formed even
$\approx 250$ Myr after initial starbursts (corresponding to 1G formation in
GC formation).  This means that 2G formation can be severely suppressed 
for $\approx 250$ Myr
after 1G formation: large reduction in the total masses of 2G stars is highly likely.
Therefore, such CCSNe formation needs to be avoided or suppressed
in any GC formation scenarios based
on self-enrichment by AGB stars. Given that both SNIa and delayed CCSNe originate from binary
stellar populations,  a key question is whether delayed CCSNe can be really avoided or suppressed while feedback effects of prompt SNIa truncate 2G formation.  This question needs to be addressed
by our future studies in a quantitative manner.

\subsection{Pristine gas from 1G PMS stars ?}
We have assumed that pristine gas required in the SBC scenario
originates only from IMCBs. We here suggest that gas from
low-mass PMS stars can be possibly mixed with AGB ejecta to
be converted into 2G stars.
Low-mass 1G stars ($m<1.5 {\rm M}_{\odot}$) are still in the PMS
phases [$10-100$] Myr after 1G formation
(e.g., Stahler \& Palla 2004). If these PMS stars with surrounding
gas disks interact violently 
with other (more massive) stars and consequently
lose the significant mass fraction of the gas disks, then
the gas can be dispersed into intracluster medium and become
pristine gas for 2G formation.
This is a very speculative idea, but, Tailo et al. (2015) have
already shown that stellar encounters within GCs can possibly
destroy the disks around 2G PMS stars: such destruction should
be possible for 1G too.

Suppose that  (i) the present-day GC has $1.2 \times 10^5 {\rm M}_{\odot}$
in its 1G stars ($ 0.25 \le m/{\rm M}_{\odot} \le 0.8$) 
and (ii) the  GC initially had  $3.6  \times 10^5 {\rm M}_{\odot}$ in
their 1G low-mass stars (i.e., lost $\approx 70$\% of the original 1G
mass via tidal stripping etc), 
how much pristine gas can be ejected from 1G PMS stars ?
The observed mass-ratios of circumstellar disks to PMS stars for $\log m \le 0.2$ 
range from $\approx 0.003$ to $\approx 0.3$ (e.g., Natta 2004). Accordingly, 
if just 5\% of the gas disks (in mass) in 1G PMS stars
with $ 0.25 \le m/{\rm M}_{\odot} \le 1.5$
can be lost by stellar encounters,
then the total mass of the pristine  gas can be
$2.7 \times 10^4 {\rm M}_{\odot}$. This is roughly 30\% of
the total mass of the present-day 2G stars with 
$0.25 \le m/{\rm M}_{\odot} \le 0.8$ 
($8 \times 10^4 {\rm M}_{\odot}$ for $f_{\rm 2g}=0.4$).
 Therefore,
the amount of this pristine gas is not negligible at all.
However, the above assumption of 5\% loss from 1G PMS stars
could be just  an overestimation.

In the above discussion, the disk around PMS stars cannot be destroyed completely
before 2G stars start to form. If the two-body dynamical 
relaxation timescales ($t_{\rm relax}$)
of GCs correspond to disk-star interaction timescales,
then most PMS disks of low-mass stars  should survive at least $\approx 10^8$ yr
owing to $t_{\rm relax} \approx$ several $10^8$ yr.
This disk-star interaction, however, should be investigated using
hydrodynamical simulations to discuss the survival of the disks around PMS stars.
In order to discuss this idea in our future papers,
we will  need to better quantify
the total mass lost from low-mass PMS stars
in a more quantitative manner using
realistic numerical simulations of dynamical interaction
between gas disks around PMS stars and 1G stars.

\section{Conclusions}

We have investigated the mass fractions of
2G stars ($f_{\rm 2g}$), ratios of initial GC masses
to total masses of low-mass 1G and 2G stars ($F_{\rm mb}$, i.e.,
the mass budget factors),  dilution factors
($F_{\rm dil}$),  differences in [Na/Fe] between
1G and 2G stars, and Li abundances  of 2G stars in GCs (A(Li))
using analytic and one-zone models of the 
SBC (``single-binary-composite'') scenario.
In this scenario, gaseous ejecta
from single AGB stars and IMCBs can be well mixed
to be converted into new 2G stars. 
The  fractions of binary stars ($f_{\rm b}$)
and the slopes of IMFs in 1G ($\alpha$) are the two key parameters
that can control the basic characteristics of GCs.
In order to discuss how the abovementioned physical properties 
of GCs depend on $f_{\rm b}$ and $\alpha$,
we have used the results of V12, which predicted
the mass fractions of gas ejected
from IMCBs ($f_{\rm ej}$) and  those of pristine gas in the ejecta
($f_{\rm pr}$). The principle results are described as follows. \\

(1) If $\alpha$ is smaller than 1.7 for 
$m_{\rm u, 2g}=1 {\rm M}_{\odot}$ and  $f_{\rm b}=0.7$, then
$f_{\rm 2G}$ can be as large as the observed value of $\approx 0.4$ for
the typical Galactic GCs with $\log (M_{\rm gc}/{\rm M}_{\odot})=5.2$.
However, $f_{\rm 2g}$ can be always lower than 0.4 for
$m_{\rm u, 2g}=8 {\rm M}_{\odot}$ for a reasonable range of $\alpha$.
This suggests that for such IMFs with larger $m_{\rm u, 2g}$,
1G stars  need to be more efficiently removed from GCs
during their long-term evolution to reduce $f_{\rm 2g}$ to the observed level.
Irrespectively of $f_{\rm b}$ and $m_{\rm u, 2g}$, $f_{\rm 2g}$ is higher 
for lower $\alpha$. \\

(2) The original GC masses should be by a factor of $2-7$ larger than
the present-day total GC (1G+2G) 
 masses with $0.25 \le m/{\rm M}_{\odot} \le 0.8$
for $1.5 \le \alpha \le 2.5$, and this result does not depend on $f_{\rm b}$.
$F_{\rm mb}$ is larger for smaller $\alpha$ (more top-heavy IMFs),
however,
it is still $\approx 7$ for $\alpha=1.7$ required for $f_{\rm 2g}=0.4$.
The derived $F_{\rm mb}$ is not so large (i.e., not like $F_{\rm mb}>10$),
which implies that there is no mass budget problem in the SBC scenario.
It should be noted, however, that all gas from single AGB and IMCBs
is assumed to be  converted into 2G stars in the present study. Accordingly,
the derived $F_{\rm mb}$ is the lower limit. \\

(3) Since the total amount of pristine gas from IMCBs depends strongly
on $f_{\rm b}$,  $F_{\rm dil}$ depends strongly on $f_{\rm b}$ too.
As a results of this,  [Na/Fe] differences  between 1G and 2G stars 
($\Delta$[Na/Fe])
depend on $f_{\rm b}$ such that $\Delta$[Na/Fe] is smaller for larger
$f_{\rm b}$ for a given $f_{\rm ej}$ and $f_{\rm pr}$.
For a reasonable range of model parameters (e.g., $\alpha$ and $f_{\rm b}$),
$\Delta$[Na/Fe] can be consistent with observations. Therefore,
the ``right'' amount of pristine gas required to dilute
the AGB ejecta to the ``right'' degree
is naturally explained in the SBC scenario. \\

(4) A(Li) in 2G stars can be as high as those in 1G
(e.g., A(Li) $\approx 2.2$ for low-metallicity GCs),
if 2G formation occurs only after  Li-free gas from IMCBs is 
mixed well  with Li-rich
AGB ejecta.
2G stars can have lower A(Li), if
they are formed early when 
Li-free gas from IMCBs is not yet fully mixed with
an enough amount of polluted gas from single AGB stars.
A(Li) of 2G stars depends  on $f_{\rm b}$,
though other parameters have no/little effects on A(Li). \\

(5) Formation histories of 2G stars show two weak peaks if $f_{\rm b}$
is higher ($>0.5$) for $M_{\rm g, th}=0$, mainly because there are two peaks
in the time evolution of gas ejection
from IMCBs predicted from V12. 
The models with a threshold gas mass/density  show discrete multiple peaks of 
star formation, which implies that the observed discrete populations
in GCs could be due to a threshold gas mass or density for the formation
of 2G stars in dense stellar systems.
Since no accretion of ISM from GC host dwarf galaxies is assumed in
the present study, formation histories of 2G stars is determined by
$f_{\rm b}$ for a given IMF. \\

(6) Although the formation of 2G stars with 
low A(Li)  could naturally explain
the observed low A(Li) of 2G stars in some GCs (e.g., NGC 6752;
Pasquini et al. 2005),
it should be avoided for most GCs with MPs: it can be a potentially
serious problem in the scenario. The hypothetical threshold gas mass or density
for 2G formation can alleviate this potential problem, because it can allow
2G formation to occur only after both single AGB stars and IMCBs have ejected
a significant amount of gas (thus after they have mixed together). \\

(7) The four dilution problems  (i.e.,  timing,  amount, 
metallicity,  and truncation) in the standard
AGB scenario
are not serious in the SBC scenario.
However, CCSNe should be severely suppressed during 2G formation
that can last for an order of $\approx 10^8$ yr.
Multiple dynamical interactions between 1G stars and gas within GCs
can lower the maximum possible mass for individual star formation
(i.e., $m_{\rm u, 2g}<8 {\rm M}_{\odot}$) so that CCSNe cannot occur.
This avoidance of supernovae is a fundamental  requirement for the
scenario to be viable. Therefore, more sophisticated hydrodynamical
simulations  need to be done to understand what physical processes
determine  $m_{\rm u, 2g}$.  \\

(8) IMCBs can provide an enough amount of pristine gas for the formation of
2G stars in GCs. However, they can also possibly provide a significant number of delayed
CCSNe, which are likely to suppress star formation within GCs  even 250 Myr after
1G formation. Therefore, incorporating IMCBs in GC formation models could be a 
double-edge sword.

\section{Acknowledgment}
I am (Kenji Bekki, KB)  grateful to Dany Vanbeveren for his providing numerical data
for the total gas mass ejected from IMCBs within GCs.
KB is also  grateful to the referee  for  constructive and
useful comments that improved this paper.

\section{DATA AVAILABILITY}
The data used in this paper (outputs from computer simulations)
will be shared on reasonable request
to the corresponding author.


\begin{thebibliography}{}

\bibitem[]{}
Bastian, N., et al., 2005, A\&A, 443, 79

\bibitem[]{}
Bastian, N., Lardo, C., 2018, ARA\&A, 56, 83 (BL18)

\bibitem[]{}
Baumgardt, H., et al., 2019, MNRAS, 482, 5138

\bibitem[]{}
Bekki, K., 2017, MNRAS, 469, 2933 (B17)

\bibitem[]{}
Bekki, K., 2019, A\&A, 622, 53 (B19a)

\bibitem[]{}
Bekki, K., 2019, MNRAS, 486, 2570 (B19b)

\bibitem[]{}
Bekki, K., 2022, in preparation

\bibitem[]{}
Bekki, K., et al. 2002, ApJ, 602, 730

\bibitem[]{}
Bekki, K.,  Norris, J. E., 2006, ApJ, 637, L109 


\bibitem[]{}
Bekki, K. Campbell, S. W. Lattanzio, J. C., \&  Norris, J. E.
2007, MNRAS, 377, 335


\bibitem[]{}
Bekki, K., Jerabkova, T., Kroupa, P.,  2017, MNRAS,  471, 2242 


\bibitem[]{}
Calura, F., et al., 2019, MNRAS, 489, 3269
 

\bibitem[]{}
Carretta, E.,  Bragaglia, A., Gratton, R. G.,
Recio-Blanco, A., Lucatello, S., D'Orazi, V., \&  Cassisi, S.
2010, A\&A, 516, 55

\bibitem[]{}
Carretta, E., et al., 2012, ApJ, 750, L14


\bibitem[]{}
D'Antona, F.,  et al., 2012, MNRAS, 426, 1710


\bibitem[]{}
De Donder, E., Vanbeveren, D., 2004, New Astronomy, 48, 861

\bibitem[]{}
 De Marchi, G., Paresce, F.,  Pulone, L.,
2007, ApJ, 656, L65

\bibitem[]{}
de Mink, S. E., et al. 2009, A\&A, 507, L1

\bibitem[]{}
D'Ercole, A., Vesperini, E., D'Antona, F., McMillan, S. L. W.,
\& Recchi, S. 2008, MNRAS, 391, 825 (D08)

\bibitem[]{}
D'Ercole, A., D'Antona, F., Ventura, P.,
Vesperini, E., \&  McMillan, S. L. W.
2010, MNRAS,
 407, 854

\bibitem[]{}
D'Ercole, A., et al., 2011, MNRAS, 415, 1304


\bibitem[]{}
Fenner, Y., Campbell, S. W., Karakas, A. I.,
Lattanzio, J. C., Gibson, B. K., 2004, MNRAS, 353, 789


\bibitem[]{}
Ferraro, F. R., et al. 2009, Nature, 462, 483

\bibitem[]{}
Gratton, R. G., Carretta, E., 2010, A\&A, 521, 54
 
\bibitem[]{}
Gratton, R. G., et al., 2019, A\&ARv, 27, 8 


\bibitem[]{}
Greggio, L., Renzini, A., 2011, Stellar populations. A user guide from low
to high redshift

\bibitem[]{}
Hong, J., et al., 2015, MNRAS, 449, 629

\bibitem[]{}
Johnson, C. I., et al., 2019, MNRAS, 485, 4311

\bibitem[]{}
Karakas, A., ,  Lattanzio, J., 2014, PASA, 31, 30

\bibitem[]{}
Kim, J. J., Lee, Y-W., 2018, 869, 35

\bibitem[]{}
Kouwenhoven, M. B. N., 2007, A\&A, 474, 77 (K07)

\bibitem[]{}
Kroupa, P., Jerabkova, T., 2018, in preprint (axXiv:1806.10605)

\bibitem[]{}
Krause, M. G. H., et al. 2020, SSRv, 216, 64



\bibitem[]{}
Lacchin, E., Calura, F., Vesperini, E., 2021, MNRAS, 506, 5951


\bibitem[]{}
McKenzie, M., Bekki, K., 2018, MNRAS, 479, 3126

\bibitem[]{}
McKenzie, M., Bekki, K., 2021, MNRAS, 500, 4578 (MB21)



\bibitem[]{}
Marino, A. F., et al.,   2018, ApJ, 859, 81

\bibitem[]{}
Marino, A. F., et al., 2015, MNRAS, 450, 815

\bibitem[]{}
Marks, M., 2012, MNRAS, 422, 2246 (M12)

\bibitem[]{}
Milone, A. P., Marino, A., F., 2022, preprint (arXiv.2206.10564), (MM22)

\bibitem[]{}
Natta, A., 2004, ASPC, 324, 20

\bibitem[]{}
Naiman, J. P., et al., 2020, MNRAS, 491, 4602


\bibitem[]{}
Pasquini, L., Bonifacio, P., Molaro, P., et al., 2005, A\&A, 441, 549

\bibitem[]{}
Piotto, G. et al. 2005, ApJ, 621, 777

\bibitem[]{}
Prantzos, N., \& Charbonnel, C. 2006, A\&A, 458, 135


\bibitem[]{}
Renzini, A. Buzzoni, A., 1986,  	
in  Spectral evolution of galaxies,
(Dordrecht, D. Reidel  Publishing Co.),  p.195

\bibitem[]{}
Renzini, A., et al. 2015, MNRAS, 454, 4197

\bibitem[]{}
Renzini, A., et al. 2022, MNRAS, 513, 2111

\bibitem[]{}
Roederer, I. U., 2011, ApJ, 732, L17


\bibitem[]{}
Sollima, A., Ferraro, F. R., Bellazzini, M., Origlia, L.,
Straniero, O., \&  Pancino, E.
2007, ApJ, 654, 915

\bibitem[]{}
Stahler, S. W., Palla, F., 2004, The Formation of stars,  Wiley-VCH

\bibitem[]{}
Tailo, M., et al. 2015, Nat, 523, 318


Vanbeveren, D., et al., 2012, A\&A, 543, 4

\bibitem[]{}
Ventura, P., et al. 2001, ApJ, 550, L65

\bibitem[]{}
Vesperini, E., McMillan, S. L. W.,  D'Antona, F., D'Ercole, A.,
2010, ApJ, 718, L112

\bibitem[]{}
Webb, J. J.,  Leigh, N. W. C., 2015, MNRAS, 453, 3278

\bibitem[]{}
Williams, M.i L., et al., 2021,  MNRAS, 512, 4086


\bibitem[]{}
Yaghoobi, A., et al., 2022, MNRAS, 510, 4330

\bibitem[]{}
Yong, D., Grundahl, F., D'Antona, F., Karakas, A. I.,
Lattanzio, J. C.,  Norris, J. E.,
2009, ApJ, 695, L62


\end{thebibliography}
\end{document}